\documentclass[aps,prb,twocolumn,nopacs,amssymb]{revtex4}
\usepackage[dvips]{graphicx}
\usepackage{dcolumn}
\usepackage{bm}

\def\gs{\gtrsim}
\def\ls{\lesssim}
\def\be{\begin{equation}}
\def\en{\end{equation}}    
\def\gs{\gtrsim}
\def\ls{\lesssim}

\newcommand{\bi}[1]{\mbox{\boldmath$#1$}}
\newcommand{\av}[1]{\langle{#1}\rangle}

\def\p{\partial}
\def\bea{\begin{eqnarray}}
\def\ena{\end{eqnarray}}

\def\asig{\stackrel{\leftrightarrow}{\sigma}}

\begin{document}
\draft
\bibliographystyle{prsty}
\title{Hydrodynamics in  bridging and aggregation of  
two  colloidal particles\\
 in a near-critical binary mixture}
\author{Shunsuke  Yabunaka$^1$, 
Ryuichi Okamoto$^2$,   and Akira Onuki$^3$}
\address{$^1$Yukawa Institute for Theoretical Physics,  Kyoto University, 
Kyoto 606-8502 Kyoto, Japan\\
$^2$Department of Chemistry, Tokyo Metropolitan University, 
Hachioji, Tokyo 192-0397, Japan\\
$^3$Department of Physics, Kyoto University, Kyoto 606-8502, Japan}
\date{\today}

\begin{abstract} 
We investigate    bridging 
and aggregation of  two  colloidal particles 
in a near-critical binary mixture  
 when the  fluid far from the particles 
is outside  the coexistence (CX) curve and is rich in 
the component disfavored by the colloid surfaces.  
In such situations, the adsorption-induced interaction is 
enhanced, leading to bridging and aggregation of the particles.  
We realize bridging firstly by changing   the temperature 
 with a fixed interparticle 
separation  and secondly by letting the two particles  aggregate. 
 The interparticle attractive force 
dramatically increases upon bridging.  
The  dynamics is governed  by hydrodynamic  flow    
around the colloid surfaces.  In aggregation, 
the adsorption layers move with the particles 
and squeezing occurs  at narrow  separation. 
These results suggest relevance of bridging in the reversible colloid 
aggregation observed so far. 
 We use the  local functional theory 
$[$J. Chem. Phys. {\bf 136}, 114704 (2012)$]$ 
to take into account the renormalization effect 
and the simulation method $[$Phys. Rev. Lett. {\bf 85}, 1338 (2000)$]$ 
to calculate the hydrodynamic flow around the  colloidal particles.

 \end{abstract}

\maketitle

\pagestyle{empty}

\section{Introduction}

Colloidal suspensions  exhibit various intriguing   effects 
in a  mixture solvent. For example, interplay between 
wetting  and phase separation has been studied extensively  
in experiments  \cite{Tanakareview,Tanaka1,Mah,Amis} 
and  simulations \cite{Balzas,La,Araki}. Much attention has 
also been paid to  a  solvent-mediated  attractive interaction  
 among  solid surfaces due to  adsorption-induced  
concentration  disturbances\cite{Fisher,Krech,Two,Ga,Gam,Evans-Hop}, 
which is  enhanced 
with increasing the correlation length $\xi$ near the criticality. 
In particular,  reversible  colloid aggregation  has been 
observed at off-critical 
 compositions  outside the  coexistence curve (CX)
\cite{Be1,Be2,Be3,Be4,Bey,Be5,Maher,Guo,Bonn}, 
where the solvent far from the particles is  
rich  in the component disfavored 
by the colloid surfaces (near the disfavored 
branch of CX). It occurs while the particle radius $a$ still much exceeds $\xi$. Remarkably,  the  face-to-face  separation distance  $\ell$ 
 is of order $a$ and is much longer than 
$\xi$ in these aggregates  \cite{Be1,Be2,Be3,Be4,Be5,Bey}.
As a result,  the   aggregates fragment and 
the   particles  redisperse  upon a 
reduction of adsorption caused by  a small temperature 
change.  Guo {\it et al.}\cite{Guo} furthermore observed 
 liquid, fcc crystal, and glass  phases 
of the aggregated particles  also for 
  $\ell\sim a \gg  \xi$. We mention a number of  theoretical papers   
on this near-critical aggregation 
\cite{Slu,Lowen,Netz,Kaler,Peter,Die,Mohry,Edison}.
In  non-critical solvents, on the other hand,   
the particles often stick and form  fractal aggregates   
due to the  attractive van der Waals  interaction   
\cite{Russel}.

The adsorption-induced  interaction 
 decays  exponentially as $\exp(-\ell/\xi)$ for $\ell \gs \xi$ 
if the solvent is at the critical composition 
 without phase separation \cite{Fisher,Two,Gam},  
as was measured directly \cite{Nature2008}. 
Some authors  \cite{Bonn,Die} used  this  exponential expression 
to explain the observed aggregation. 
In their  approach, close contacts of the particles 
  in the range  $\ell\ls \xi$  are needed  for aggregation, 
since the   factor $\exp(-\ell/\xi)$ is nearly zero for  $\ell\gg \xi$. 
In contrast, the van der Waals two-body potential\cite{Russel,Ross} 
is long-ranged for $\ell>a$ 
and  grows  as $-A_{\rm H}a/12\ell$ for $\ell<a$, 
where $A_{\rm H}$ is the Hamaker constant. 
In recent    theories\cite{Peter,Mohry}, the colloidal particles 
were  assumed  to form   one-component  
systems interacting  via   
 an attractive  solvent-mediated interaction   potential 
($\propto e^{-r/\xi}$)  and a repulsive electrostatic potential, 
where the particle separation $r$ 
is smaller than $\xi$ in aggregation.
Very recently, Edison {\it et al.} \cite{Edison} has 
examined  phase behavior of 
a three-component model on a 2D lattice 
to obtain  aggregation of the third component.

As a related problem, 
much attention has also been paid 
to the phase behavior of fluids between closely separated 
walls \cite{Evansreview,Gelb,Binder}, where 
narrow regions may be  filled with the phase favored by  the  
walls or may hold some fraction of 
the disfavored  phase.  As a result, there can be   
a first-order phase transition between these two states, 
called   capillary condensation, when the fluid in the reservoir is 
in the disfavored phase.  For fluids between parallel plates, 
there appears a first-order capillary condensation line 
outside (bulk) CX in the plane of  the temperature  $T$    
and the reservoir chemical potential  $\mu_\infty$ 
(which corresponds to a magnetic field for magnetic systems)
\cite{Evansreview,Gelb,Binder,Fisher,Nakanishi,Evans-Marconi}. 
 Analogously, between  large particles (or between a large 
particle and a wall),  a bridging domain of the 
favored phase can  appear, 
when they are surrounded by  the disfavored phase 
\cite{Gam,Yeomans,Bauer,Higashi,Vino,Seville,Butt,bubble}. 
This effect is relevant in the physics of wet granular matter\cite{Gra}. 
Recently, we calculated phase diagrams 
of   capillary condensation 
 \cite{Oka-c}  and 
  bridging \cite{Oka-b}. 
We  stress that  the solvent-mediated 
interaction can be  much    enhanced near these  transitions  
 even in pretransitional states 
(before local  phase separation)
\cite{Oka-c,Oka-b,Two,Anna,Evans-Hop}. 
We also  examined the  dynamics of capillary condensation 
\cite{Yabu}.

In this paper, we demonstrate 
that the capillary force  
mediated  by a bridging domain  can 
give rise to the observed reversible  aggregation. 
Let  two colloidal particles  
be bridged by a  columnar   domain of the favored phase, where 
 the column  radius is of the order  of 
the particle radius  $a$. Then, in terms of 
the surface   tension $\sigma$, the interface  free energy 
is of order $2\pi\sigma a \ell$ 
and its derivative with respect to $\ell$ 
yields   the    capillary force 
 of order   $2\pi \sigma a$ 
between the particles \cite{Oka-b,Seville,Butt}. 
 Such bridging can  occur  even for large $\ell$ of order $a$, 
so it can be relevant to the reversible aggregation.  
 We  then need to investigate its dynamics  
to demonstrate its occurrence  for $\ell\sim a\gg  \xi$ 
near the disfavored branch of  CX. It is a local 
phase separation process governed by the hydrodynamics,   
where   squeezing also takes place as the particles approach.  
 We note that 
the  colloid hydrodynamics  coupled  with the concentration  
  has been studied 
in various situations  \cite{Tanakareview,Araki,Furukawa}  
using  the model H equations \cite{Halperin,Kawasaki-Ohta,Onukibook}.

As a theoretical  method, 
we use   the renormalized local functional 
theory \cite{Oka-c,Oka-b,Fisher-Yang,Up,Upton} 
to account for the near-critical fluctuation effect. 
It is combined with  the fluid-particle dynamics  (FPD) simulation method  
 \cite{Araki,Tanakareview,Tanaka2,Fujitani}  to 
calculate  the  flow around the  particles. In our case,  
 the adsorption layers are thickened 
 near the criticality \cite{Jasnow,Liu,Lawreview}, which 
strongly  affect  the 
flow in aggregation. We mention that 
 Furukawa {\it et al.}\cite{Furukawa} 
used  the FPD method   to study  
 two-particle aggregation   at the critical composition. 
They  assumed  a small particle radius  $a$ with  $a\sim \xi$ 
in the mean-field theory, where the solvent-mediated  force  
was appreciable even for  $\ell \sim a$ because of relatively large $\xi$. 
However, if $\ell\gg \xi$, the force  virtually vanishes  at the critical 
composition.

The organization of this paper is as follows. 
In Sec.II,  we will shortly explain the FPD method  and  
the   local functional theory applied to our problem.  
In Sec.III, we will explain the  simulation method. 
In Sec.IV, we will present 
simulation results of bridging dynamics.

\section{Theoretical Background}
{\subsection{Critical behavior }}
We consider  a near-critical binary mixture 
 with an  upper  critical solution temperature   $T_c$ 
for small reduced temperature 
  $\tau=T/T_c-1$  at a given    pressure, 
where  $T$ is above the wetting temperature $T_{\rm w}$\cite{Bey,Ross}. 
(For mixtures with a lower   critical solution temperature 
such as 2,6-lutidine and water (LW),
we should set   $\tau=1-T/T_c$.)     
 The order parameter $\psi$ 
is a  scaled concentration deviation   slightly deviating 
from its critical value.   
For  $\tau<0$, a two-phase region appears 
in the $\tau$-$\psi$ plane. The  coexistence curve (CX)  
consists of the favored and disfavored branches as   
\be 
\psi=\pm \psi_{\rm cx} = \pm b_{\rm cx}|\tau|^\beta,
\en 
 where   $ b_{\rm cx}$ is a constant. 
The  critical exponents\cite{Onukibook,Vicari,Aliu} are given by   
 $\beta\cong 0.325$,  $\nu\cong 0.630$, and ${\hat\eta}
\cong 0.032$. 
The correlation length $\xi$ depends on 
$\tau$ and $\psi^2$  and  is given by 
$\xi_0\tau^{-\nu}$ for $\tau>0$ and   $\psi=0$ 
and by $\xi_0'|\tau|^{-\nu}$ for $\tau<0$ on CX,   
where $\nu$ is assumed to be 
common in the two cases and 
$\xi_0$ and $\xi_0'$ are  microscopic lengths. 
Here, $\xi_0$ is  usually 
in the range $2-3$ ${\rm \AA}$ 
for low molecular-weight binary  mixtures. 
The  ratio $\xi_0/\xi_0'$ is a universal number 
in the renormalization group theory\cite{Onukibook} 
and its reliable estimate\cite{Aliu} is $1.9$, so we suppose 
$\xi_0'\sim 1$ ${\rm \AA}$. 

We  express the free energy density 
$f(\psi,\tau)$ 
 in accord with the scaling relations 
in the vicinity of the critical point in the $\tau$-$\psi$  plane 
including the region inside CX ($\tau<0$, $|\psi|<\psi_{\rm cx}$). 
This is needed  because $\psi$ 
changes from positive to negative around the colloidal particles 
in the presence of  strong adsorption.  To this end, we  use 
a simple  parametric form of $f(\psi,\tau)$ 
devised in our previous paper  \cite{Oka-c} (see  the appendix).

At the starting point of our theory,  
the thermal fluctuations of $\psi$ and the 
velocity field $\bi v$ with wave numbers 
 larger   than $\xi^{-1}$ have already been coarse-grained 
or renormalized. Here,  $\xi^{-1}$  is the lower cut-off wave number 
of  the renormalization effect, so  
 the  free energy density  $f(\psi,\tau) $ and 
the kinetic coefficients ($\lambda$ 
in eqn (6) and $\eta_0$ in eqn (9)) are  renormalized 
ones  depending on fractional powers of  $\xi^{-1}$.  
Then, since our  dynamic equations  have 
 renormalized coefficients,  they  can well describe 
nonequilibrium processes with spatial scales longer than $\xi$ 
even without the thermal noise  terms  
  \cite{Halperin,Kawasaki-Ohta,Onukibook}.  
We also  neglect Brownian motions of the colloidal  
particles. 

\vspace{2mm}
{\subsection{Strong adsorption near criticality }}

As the critical point is approached, the strong adsorption 
regime is eventually realized outside CX 
due to a nonvanishing  surface field $h_1$ (see eqn (4) and (5))  
\cite{Oka-c,Oka-b,Jasnow,Fisher-Yang,Lawreview,Liu}. 
Here, let $\psi$ be  positive  near the colloid surfaces.  
If  the distance $z$ from such a wall  is shorter 
than the bulk correlation length $\xi$ in the case $\xi\ll a$, 
  $\psi(z)$  behaves  as\cite{Jasnow,Oka-c}   
\be
\psi\cong A_0  (z+\ell_0)^{-\beta/\nu} \quad (0<z<\xi),  
\en 
where   $A_0$ is a constant and
  $\ell_0$ is  a microscopic length. 
The surface value of $\psi$ at $z=0$ 
satisfies $\psi_0 
\cong A_0 \ell_0^{-\beta/\nu}\gg \psi_{\rm cx}$ 
in the strong adsorption condition. 
Since  $\beta/\nu\cong 0.52$, the preferential adsorption in the 
 layer $z<\xi$  grows  as $\int_0^\xi dz \psi\sim A_0 
\xi^{1-\beta/\nu}$ per unit area, resulting in   the well-known 
critical adsorption, where the integral in the region $0<z<\ell_0$ 
is negligible and the strong adsorption limit is well-defined. 

For  $z>\xi$,  $\psi$ 
 depends on its value $\psi_\infty$ far from the wall.   
It   decays as  $\psi \sim A_0 \xi^{-\beta/\nu}
e^{-z/\xi}$ for 
$\psi_\infty=0$ at the critical composition,  
 but it  changes its sign from positive to negative 
for  $\psi_{\infty}<0$. In this paper,  
we set $\psi_{\infty}\cong -\psi_{\rm cx}$ outside CX, where 
 the thickness of the transition layer is about  $5\xi$ 
with enlarged composition  
disturbances around the particles\cite{Oka-c,Oka-b}. In accord with this 
result,   a light scattering experiment \cite{Bey} 
showed that the adsorption layers of colloidal particles in LW 
were  much thicker near the disfavored branch of CX 
than near the favored one. 

In the literature 
\cite{Krech,Two,Ga,Gam,Guo,Bonn,Nature2008,Peter,Edison,Die,Mohry}, 
the adsorption-induced interaction has 
 been called the critical  Casimir interaction. However, 
    the original (quantum)  
Casimir interaction  stems    from  the ground-state fluctuations 
of the electromagnetic field between two mirrors\cite{Kardar}.  
In  near-critical fluids,  more 
 analogous  is the  interaction arising from 
    the thermal fluctuations of $\psi$ 
 at zero-surface field $h_1=0$,  where the thermal average 
 $\av{\psi}$ is homogeneous. Notice that 
the fluctuation-induced interaction\cite{Kardar}   
is  much weaker than  the  adsorption-induced one 
 (with inhomogeneous $\av{\psi}$)\cite{Fisher,Fisher-Yang}.  
In fact,  the  Casimir amplitudes 
at $h_1=0$ (in  the Neumann boundary condition)  
  \cite{Krech1}  
are  much  smaller than those  
under   strong adsorption 
by one order of magnitude \cite{Krech,Two,Ga,Gam,Up,Upton,Oka-c}.

\subsection{Fluid-particle dynamics (FPD) 
near criticality}

 In  the FPD method 
 \cite{Tanakareview,Araki,Furukawa,Tanaka2,Fujitani},  
 the solid-liquid boundaries  are treated as  
diffuse  interfaces  with thickness $d$ much shorter than the radius  $ a$, 
 which much simplifies the calculation of colloid motions.
We introduce the colloid  shape function 
$\theta({\bi r})= \sum_k \theta_k({\bi r})$ ($k=1,2$ here) with  
\be 
\theta_k({\bi r}) =\frac{1}{2}+ \frac{1}{2}\tanh
\bigg[\frac{1}{d} (a- |{\bi r}- {\bi R}_k|)\bigg],
\en
which tends to 1 in the colloid interior and to  zero 
in the colloid exterior. 
We define $\psi$ even in the colloid interior 
and assume the  total free energy $F$ of the form
\cite{Araki,Tanaka2,Furukawa}    
\bea
&&\hspace{-5mm}F
 =  \int\hspace{-2mm}d{\bi r}\bigg\{(1-\theta )[f(\psi,\tau)
+ \frac{C}{2}|\nabla\psi|^2] 
+\theta \chi_0  (\psi-\psi_{\rm in})^2\bigg\}
\nonumber\\
&&\hspace{-1mm}-  (h_1/3d) \int d{\bi r}\psi
\sum_k |\nabla\theta_k|^2  +U({\bi R}_1,{\bi R}_2), 
\ena 
where the integration is within the cell. 
In the first line,  $f(\psi,\tau)$ is the free energy density, 
 $C$ is a weakly singular 
positive coefficient, and 
 the  term proportional to 
$\chi_0$ serves to  fix $\psi$ at 
$\psi_{\rm in}$ in the  colloid interior. 
See  the appendix for $f(\psi,\tau)$ and  $C$. 
In the previous papers\cite{Araki,Tanaka2,Furukawa}, 
the mean field form  of $f$ has been used.  
In the second  line, 
${ h}_1$ is the surface field assuming 
a large  positive number  
\cite{Ross,Oka-c,Oka-b} and  $U({\bi R}_1, {\bi R}_2)$ is an applied 
potential  acting on the  particles.  In the limit  $d/ a \to 0$,  
 the second term tends to  the usual surface integral 
$- h_1 \int dS \psi$ and use of eqn (2) yields 
\be 
h_1 \cong (\beta/\nu )C \psi_0/\ell_0,
\en 
 where $\psi_0$ is the surface value  of $\psi$. 
In the strong adsorption condition $\psi_0\gg\psi_{\rm cx}$, 
we find  $h_1\propto  \psi_0^{2\nu/\beta-1}$ 
from $C\propto \psi_0^{-{\hat \eta}\nu/\beta}$ 
(see the appendix).

Assuming    a homogeneous $\tau$, we 
use the model H dynamics 
\cite{Halperin,Kawasaki-Ohta,Onukibook}. 
The order parameter  $\psi$ is governed by      
\be 
\frac{\p}{\p t}\psi= -\nabla\cdot[
\psi{\bi v} -\lambda\nabla\mu] ,  
\en 
where $\bi v$ is the velocity field assumed to be 
incompressible ($\nabla\cdot{\bi v}=0$), $\lambda$ 
is the kinetic coefficient, 
and  $\mu=  {\delta F}/{\delta\psi}$ 
is the (generalized) chemical potential. 
The particle velocities are given by the average  interior velocity 
field,  
\be 
\frac{d}{dt} {\bi R}_k= \frac{1}{v_0}\int d{\bi r}\theta_k {\bi v},
\en 
where $v_0 = \int d{\bi r}\theta_k 
\cong {4 \pi a^3}/{3}$. Neglecting the acceleration 
(the Stokes approximation), we determine $\bi v$ from   
\be 
\psi \nabla  \mu + \frac{1}{v_0} \sum_k { \theta_k} {\bi{\cal F}}_k 
+\nabla p_1
= \nabla\cdot {\asig}_{\rm vis}  , 
\en
where ${\bi{\cal F}}_k= \p F/\p {\bi R}_k$, 
  $p_1$ is a pressure ensuring  $\nabla\cdot{\bi v}=0$, 
and $ {\asig}_{\rm vis}=\{ \sigma_{ij}\}$ is the viscous stress tensor 
of the form, 
\be
\sigma_{ij} = (\eta_0+\eta_1\theta)(\nabla_j v_i  +\nabla_iv_j),
\en 
with $\nabla_i = \p/\p x_i$ ($i=x,y,z$). In the FPD method, 
 the colloidal particles are treated as a highly viscous fluid.  
Then, 
$\eta_{0}$ is the viscosity  in the liquid 
and  $\eta_{0}+\eta_{1}$ is the viscosity in  the colloid interior 
with $\eta_1\gg \eta_0$.   If we use  eqn (6)-(9), 
  $F$  in eqn (4) monotonically decreases 
 in time ($dF/dt\le 0$) in nonequilibrium\cite{Onukibook}.

The kinetic coefficient 
  $\lambda$ in eqn  (6) is strongly enhanced due to 
the convective motions of the critical  fluctuations 
\cite{Halperin,Kawasaki-Ohta,Onukibook}. 
Because the solvent is   close to CX in our situation, we set 
$\lambda$ equal to its value  on CX as    
\be 
\lambda=    \chi_{\rm cx} D /k_BT_c, 
\en 
where   $\chi_{\rm cx}$ is the susceptibility  on CX in  eqn (A1)  
and $D$ is the diffusion constant in    the Stokes form, 
\be 
D=k_BT_c /6\pi \eta_0 \xi  
\en  
with $\xi=\xi_0'|\tau|^{-\nu}$. For $\xi \gg \xi_0$, 
this $D$ is much smaller than  microscopic diffusion constants.  
In contrast, 
the viscosity   ${\eta_0}$ in eqn (9) 
is only  weakly   singular     and  may be   treated as 
a constant  independent of $\tau$ and $\psi$ 
in near-critical fluids.

\begin{figure}[t]
\begin{center}
\includegraphics[scale=0.28, angle=0]{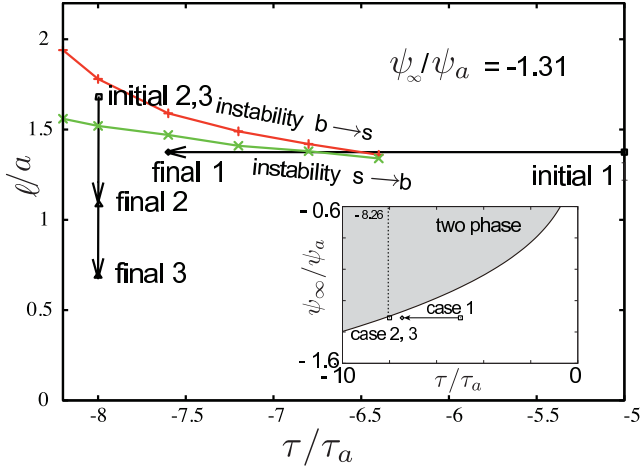}
\caption{ 
Illustration of simulation from separated to 
bridged states of two particles outside CX 
in the $\tau/\tau_a$-$\ell/a$ plane, where $\psi$ tends 
to  $\psi_\infty\cong -1.31\psi_a$ far from the particles.
If $\psi= -1.31\psi_a$,  the reduced temperature on CX  
is given by  $\tau_{\rm cx}=-8.26\tau_a$ (vertical dotted line).   
In case 1,  $\tau/\tau_a$ 
is lowered  from $-5.0$ to $-7.6$ at nearly fixed separation 
$\ell= 1.38$.  In cases 2 and 3,  $\ell/a$ changes 
from 1.69 to  1.09 and 0.69, respectively, at   $\tau/\tau_a=-8.0$. 
Below (s $\to$ b)  instability line (lower green line) 
separated states become unstable, 
while above (b $\to$ s) instability line  (upper red line)  
bridged states become unstable. 
In inset, these three cases are 
illustrated  in the $\tau/\tau_a$-$\psi_\infty/\psi_a$ plane outside CX or 
below  the two-phase region  (in gray). 
}
\end{center}
\end{figure}

{\section{Simulation background}}

In the presence of  colloidal particles with radius $a$,  
we  define characteristic values of 
$\tau$ and $\psi$  by \cite{Oka-b}
\be
\tau_a = (\xi_0/a)^{1/\nu}, \quad 
\psi_a = b'  \tau_a^{\beta},   
\en 
where the ratio  $b'/ b_{\rm cx}$ is calculated to be 
1.47 in our  scheme\cite{Oka-c}. 
We consider   negative homogeneous $\tau$ with 
considerably large $|\tau|/\tau_a= (\xi_0' a/\xi_0 \xi)^{1/\nu}\gg 1$,  
so   we treat the case  $\xi\ll a$. 
Hereafter,  we measure time $t$ in our simulation 
in units of $t_0$ defined by 
\be 
t_0= a^2/D= (a/\xi)^2 t_\xi,
\en 
where $t_\xi=6\pi \eta_0 \xi^3/k_BT$ is the  thermal 
relaxation time of the critical fluctuations \cite{Halperin,Onukibook}.
As the parameters  in eqn (3) and (4), 
 we set  $d=a/8$, $\chi_0= 783 k_BT/a^3\psi_a^2$, $\psi_{\rm in}=5\psi_{a}$,  
and ${ h}_1/3d =15.7 k_B T/a\psi_a$. 
The viscosity ratio  is  $\eta_1/\eta_0=100$.

Let us assume $\tau=-8 \tau_a$ for a  LW mixture 
slightly outside CX, 
for which $T_c=307$ K,  $\xi_0=2.5~{\rm \AA}$, and $\eta_0=2.0$ cp. 
Then, for $a=200$ nm, 
we obtain $\tau_a=1.8 \times 10^{-5}$, $\tau=-1.5\times 10^{-4}$, 
$\xi=29$ nm, $D= 3.5 \times 10^{-8}$ cm$^2$$/$s, 
$t_\xi=2.4 \times 10^{-4}$ s, and $t_0=12$ ms.  
If we increase $a$ to $1$ $\mu$m, 
$t_0$ becomes $ 1.4$ s. 


We place   two particles 
 in the middle of a cylindrical cell 
with radius $L_0=3.75a$ and height $H_0=10a$, 
so the system is in the region 
$\rho=(x^2+y^2)^{1/2}<L_0$ and $-H_0/2<z<H_0/2$. 
The particle centers are at $(0,0,\pm(\ell/2+a))$, where  $\ell$ 
is  the surface-to-surface separation distance. 
This  geometry  is  axisymmetric, so 
the time integration was performed in the 2D $\rho$-$z$ plane,  
where  the mesh size is $\Delta x= a/16$  and 
the  time interval width 
is $\Delta t= 5 \times {10}^{-6} t_0$. 
The periodic boundary condition is imposed along the 
$z$ axis, while we set $\p \psi/\p \rho=0$ 
and ${\bi v}={\bi 0}$  on the side wall $\rho=L_0$. 
Then,  the total order parameter 
$\int d{\bi r} \psi$ is conserved 
in time. As  a result,  the value of 
$\psi$ away from the particles, written as 
  $\psi_\infty$, 
exhibits a   slight decrease of order $0.02 \psi_a$  
  after bridging. 
We also find that $|\psi-\psi_{\rm in}|$ 
 in the colloid interior remains 
smaller than $0.01 \psi_a$ for our choice of $\chi_0$ and $\psi_{\rm in}$, 
which assures the validity of our model.

Even  in the axisymmetric geometry, integration of eqs (6) and (7) 
under eqn (8) is  time-consuming. 
However, if we fix the particle positions, we may efficiently seek 
the equilibrium profiles of $\psi$ 
from the relaxation equation\cite{Yabu}, 
\be 
\frac{\p}{\p t}\psi= -\mu+\mu_\infty,  
\en 
where $\mu=\delta F/\delta\psi$ 
with $\mu_\infty$ being its value  far from the particles. 
Here, $\bi v$ is not coupled. 
At long times, the stationary solution satisfies  $\mu=\mu_\infty$.

\vspace{4mm}
\section{Bridging and aggregation dynamics}
 {\subsection{ Situations of bridging}}

In Fig.1, we illustrate  how we 
performed our simulation outside CX. 
We  initially  set   $\psi_\infty=-1.31\psi_a$ 
far from the particles. 
For this concentration, the reduced temperature on CX  
is given by  $\tau_{\rm cx}=-8.26\tau_a$ from eqn (1), 
so we set $\tau> \tau_{\rm cx}$.  
 We   plot the  instability line  from 
separated to bridged states (s $\to$ b) 
and that from  bridged to separated states (b $\to$ s).  
To determine these two instability lines, 
we integrated the  relaxation  equation (14) 
 at fixed particle positions 
for various $\ell$ and $\tau$. 
 We detected  growth of small 
disturbances when  we crossed   the former (latter) line 
by  gradually decreasing  (increasing)  $\ell$ from 
separated (bridged) states.
Between these lines, both separated and  bridged states 
 remained   stationary.   
Furthermore, between them, there is   a first-order transition line,  
on which the free energy assumes the same value for 
these two states\cite{Oka-b}.

In  our simulation, even if $\ell\gg \xi$, 
the adsorption-induced force is 
significant  from the beginning. 
Here,   it is long-ranged,  decaying 
as $\propto e^{-\ell/\xi}$ 
only for $\ell\gs a$,  
due to  the expanded adsorption layers  
 close to the disfavored branch of CX (see Sec.IIA). 
See  such examples in Fig.5 in  our  paper \cite{Oka-b}.     
Indeed, if the initial attractive force 
 is equated to   $k_BT A a\xi^{-2} e^{-\ell/\xi}$ for 
case 1 in Fig.1, 
we have $A=449$, while $A\sim 4$ at the critical composition 
\cite{Die,Oka-b}.

In real experiments,  charges usually appear on the colloid surfaces 
in aqueous mixtures\cite{Be1,Be2,Be3,Be4,Maher,Guo,Bonn,Nature2008}. 
 Supposing  weak ionization, 
 we neglect the  effect of charges 
on  the adsorption-induced 
interaction. However, we will assume a charge-induced repulsive 
interaction effective 
at short separation  (in eqn (19) below )\cite{Russel}.
We also neglect the van der Waals interaction\cite{Russel,Ross} 
in our simulation.  We note that  Bonn {\it et al.}\cite{Bonn} 
observed colloid  aggregation 
in  refractive-index-matched  
systems, where the van der Waals interaction was suppressed.

\begin{figure}
\begin{center}
\includegraphics[scale=0.31, angle=0]{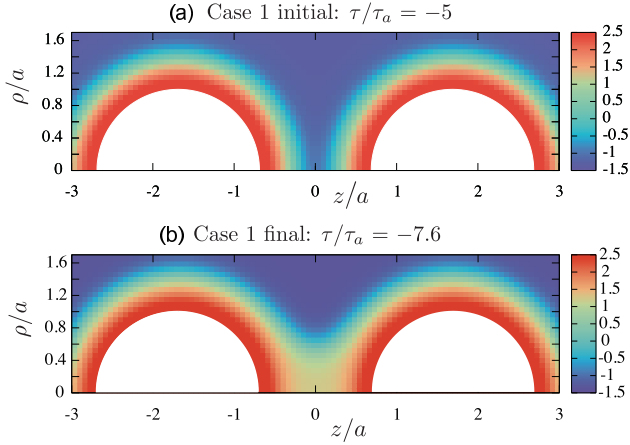}
\caption{ 
Profiles of $\psi(\rho,z,t)/\psi_a$ 
in (a) the initial state   and (b) the final state 
 in the $\rho$-$z$ plane (in gradation)  in case 1, 
where $\tau/\tau_a$  is lowered from $-5$ to $-7.6$  
at  fixed $\psi_\infty$ and $\ell$  in strong adsorption near criticality.   
}
\end{center}
\end{figure}

\begin{figure}
\begin{center}
\includegraphics[scale=0.195, angle=0]{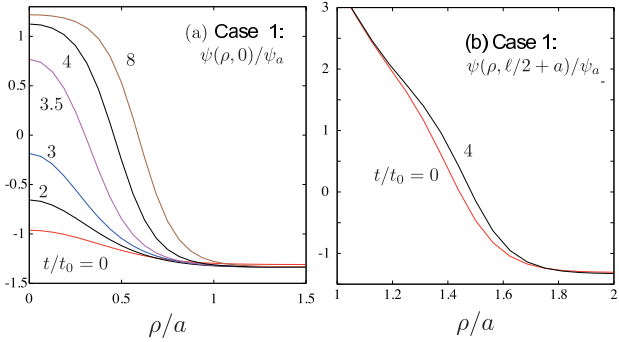}
\caption{Time evolution of 
$\psi(\rho,z,t)/\psi_a$ vs $\rho/a$ in case 1. 
(a) Profiles   on the midplane ($z=0$) 
at $t/t_0=0, 2, 3, 3.5, 4,$ and  8,  where a bridging domain appears. 
(b) Those  on the plane containing the   particle center ($z=\ell/2+a$) 
at $t/t_0=0$ and $4$, where the adsorption layer is nearly stationary.}
\end{center}
\end{figure}

\begin{figure}
\begin{center}
\includegraphics[scale=0.36, angle=0]{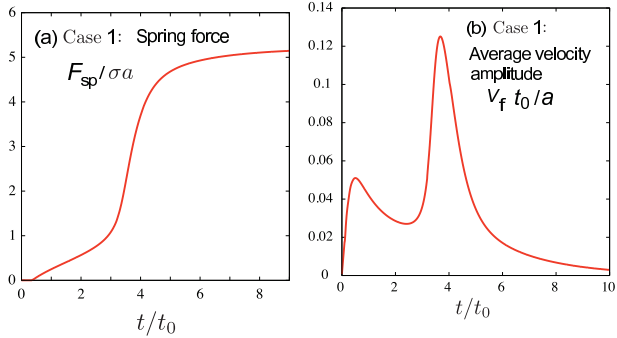}
\caption{  (a) Spring force $F_{\rm sp}(t)$  in eqn (16) 
divided by $\sigma a$ vs $t/t_0$,  which increases abruptly with bridging. 
(b) Averaged velocity 
amplitude $v_{\rm f}(t)$ in eqn (18) 
divided by  $a/t_0=D/a$ vs $t/t_0$,   exhibiting    two peaks at  
$t/t_0= 0.5$ and $3.7$. Instability growth is  slow initially 
 ($t/t_0<2$) and is accelerated  on bridge formation 
 ($t/t_0\sim 4$).
}
\end{center}
\end{figure}

{\subsection{Bridging after temperature quenching 
at fixed $\ell$}}

In case 1 simulation, at $t=0$,  we lowered  $\tau$ 
 from $-5\tau_a$ ($t\le 0$)   to $-7.6\tau_a$ ($t> 0$)  
across the (s $\to$ b) instability line,   where 
$\psi_\infty\cong -1.31\psi_a$ outside CX. 
For $t<0$, we realized the equilibrium at fixed particle positions 
as a stationary solution of eqn (14).   
The correlation length $\xi$  in the bulk 
is  $ 0.09 a$ in the final state.  
Supposing optical tweezers\cite{Yada,Kimura}, 
we choose the potential $U$ in eqn (4)  as  
\be
U =  \frac{1}{2}K
\sum_{k=1,2}(Z_{k}- Z_{k}^0)^{2}, 
\en
where $Z_1$ and $Z_2$ are the $z$ coordinates 
of the particle centers and  $Z_1^0$ and $Z_2^0$ 
are their initial values. The spring 
constant $K$ is set equal  to   $5300 k_BT/a^2$, which 
depends on $a$ and is  
$220$ pN$/\mu$m for $a=1$ $\mu$m. 
Then, the displacement $Z_{1}^0- Z_{1}$ 
is  $0.0085a$ after bridging, 
where  the separation $\ell$ is shortened 
from its initial length $1.38a$  by $1.2\%$. 
 The time-dependent  spring force  is  given by 
\be 
F_{\rm sp}(t) = 
K [Z_{1}^0- Z_{1}(t)]= K [Z_{2}(t)- Z_{2}^0], 
\en 
which  is balanced with the force 
on the particles from the fluid 
in the Stokes approximation\cite{Oka-b}.

In Fig.2, we give the initial and final  profiles 
of  $\psi(\rho,z)$ in gradation. 
We can see thick adsorption layers covering the colloid surfaces 
due to large ${ h}_1$ in eqn (4) and (5)  
and a bridging region with $\psi\cong \psi_{\rm cx}$ between the particles 
in the final state. In contrast, 
 if the system would be    quenched  slightly  inside CX, 
 thick  wetting layers with sharp interfaces would  cover the 
whole particle surfaces\cite{Araki}.

In Fig.3, we display  time evolution of 
$\psi(\rho,z,t)$ vs $\rho$   on (a) the midplane ($z=0$) 
and  (b) the plane containing the upper 
 particle center ($z=\ell/2+a$) far from the other particle.  
In (a),   $\psi(\rho, 0,t)$   increases 
between the particles. Its  growth 
is rapid  in the  time range  $3<t/t_0<4$.  
For $t>4$, it exhibits   a flat profile 
in  the middle region $\rho/a\ls 0.2$ 
and  an  interface becomes well-defined. 
In (b),  $\psi(\rho, \ell/2+a,t)$ 
exhibits    strong adsorption 
behavior, which  only slightly depends  on $t$.

In Fig.4(a), we plot the spring force $F_{\rm sp}(t)$ in eqn (16). 
 It increases   rather slowly for  $t/t_0 < 3$,  
but  abruptly for  $3< t/t_0 < 4$ 
up to   the capillary force\cite{Oka-b,Seville,Butt}, 
\be 
F_{\rm sp}\sim 2\pi \sigma a\sim k_BT a/\xi^2 , 
\en  
 where $\sigma$ is the surface tension in eqn (A3).  
In   Sec.I, the capillary force  has already been  discussed.  
See  the appendix of our previous 
paper for its more systematic derivation\cite{Oka-b}.  
This sharp increase of $F_{\rm sp}(t)$ is  
simultaneous  with the interface  formation (see   Fig.3(a)). 
Remarkably,  $F_{\rm sp}(t)$ becomes  
of order $\sigma a\sim 0.1 k_BT a/\xi^2$    before 
the bridging transition ($t<3t_0$). 
As noted at the beginning of this section, 
 the adsorption-induced force is   long-ranged 
in  pretransitional 
states  close to the bridging transition.   

We are interested in  the fluid 
velocity outside the particles.  
In Fig.4(b), we thus show the average  velocity amplitude $ v_{\rm f}(t)$ 
 in the fluid region defined by  
\be 
 v_{\rm f}(t)^2 =  \int d{\bi r}(1-\theta) v^2/ (4\pi a^3/3),
\en 
where $1-\theta$ tends to 1 outside the particles. 
We can see that   $v_{\rm f}(t)$ has a small peak at $t/t_0 =0.5$, 
 a minimum at $t/t_0 =2.4$, and a large sharp 
peak at $t/t_0 =3.7$.  In the initial stage $t/t_0 <0.5$, 
the adsorption layers of the two particles 
weakly merge without domain formation, causing  
 a weak convective flow.  This initial motion 
decays for $t/t_0 >0.5$. However,   an   unstable  mode 
emerges  giving rise to an increase in  $\psi$ 
  from the center line $\rho=0$.
It is slowly growing and not noticeable   for $t/t_0 < 2$, 
but it grows  abruptly   in the  late stage 
 $t/t_0 > 3$ with outward expansion of 
   an  interface  from the center line.  
This final  acceleration of the growth is obviously 
 due to the hydrodynamic transport.

\begin{figure}
\begin{center}
\includegraphics[scale=0.31]{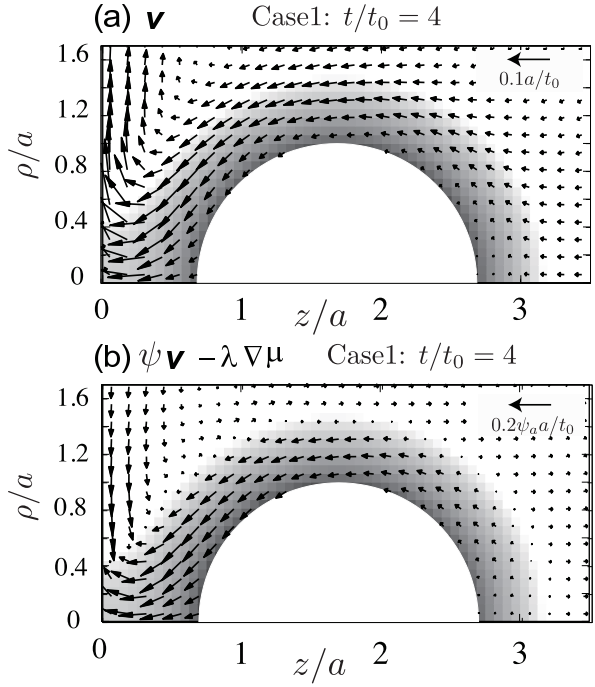}
\caption{ (a) Velocity field $ {\bi v}$ 
 and (b) concentration  flux 
$\psi{\bi v}-\lambda\nabla\mu$ 
at  $t/t_0=4.0$ in case 1.  Arrow lengths 
are  written according to standard  
lengths (a) $0.1 a/t_0$ and (b) $0.2 \psi_a a/t_0$. 
Flow is significant around the particle in the adsorption 
region   $\psi>0$ (in gray) and 
close to  the midplane.  }
\end{center}
\end{figure}

\begin{figure}
\begin{center}
\includegraphics[scale=0.4, angle=0]{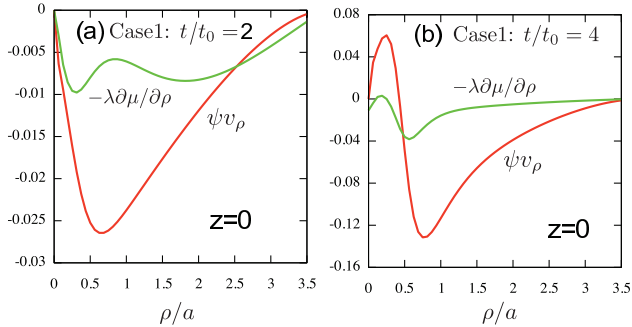}
\caption{ 
Convective part   $\psi {v_\rho}$ 
and diffusive part  $-\lambda \p \mu/\p \rho$ 
in the concentration flux vs $\rho/a$ 
in the midplane ($z=0$) 
 at (a) $t/t_0=2$   and (b) $4$  in case 1.    
The former is mostly larger than the latter by one order of 
magnitude. 
The vertical axes are written in units of $\psi_a a/t_0=D\psi_a/a$.
}
\end{center}
\end{figure}

In Fig.5, we display the profiles 
of the velocity field $ \bi v$ in (a) and the concentration flux 
 $ \psi   {\bi{v}}-\lambda \nabla \mu$ in (b)    at   $t/t_0=4$. 
Here, the vectors   $\bi v$ and $\nabla\mu$ 
are  expressed as 
$
{\bi v}= v_\rho {\bi e}_\rho +v_z {\bi e}_z$ and 
$ 
\nabla\mu= (\p\mu/\p \rho){\bi e}_\rho +(\p\mu/\p z) {\bi e}_z  
$, 
where ${\bi e}_\rho= (x/\rho,y/\rho,0)$ 
and  ${\bi e}_z = (0,0,1)$. At   $t/t_0=4$,  $v_{\rm f}$ is large in Fig.4(b). 
The maximum velocity is of order $0.1 a/t_0= 0.1 D/a$.  
We can see that   $\bi v$ (nearly) vanishes on the colloid surface, 
while $ \psi   {\bi{v}}-\lambda \nabla \mu$ is significant 
within the adsorption layer with $\psi>0$. 
The flow  is then directed outward 
on the midplane due to  the incompressibility condition  
$\nabla\cdot{\bi v}=0$. 
This squeezing flow  is blocked at 
 the side wall at $\rho=3.75a$, but 
the flow in our  case 
is appreciable only for a short time interval  and 
the boundary disturbance 
 should little  affect the bridging dynamics. 
For LW, the  typical velocity 
is estimated as 
$0.1 a/t_0= 1.7\times 10^{-4}$ cm$/$s for $a=200$ nm.

In Fig.6, we  give the $\rho$-components of the 
convective flux and the diffusive flux, 
$\psi {v}_\rho$  and $-\lambda \p \mu/\p \rho$,  separately, 
 on the midplane $z=0$ at $t/t_0=2$ in (a) and  4 in (b).
The velocity amplitude is one order of magnitude larger 
in (b) than in (a). Squeezing can be seen in the region $\rho/a<0.5$ in (b). 
 We  recognize that 
$\psi {v}_\rho$  is mostly 
larger than  $-\lambda {\nabla_\rho} \mu$ by one order of magnitude,  
which demonstrates  relevance of the hydrodynamic flow in bridging.

In their simulation, 
Araki and Tanaka \cite{Araki}  
pinned  two particles in a binary mixture 
and    quenched  
 their  system inside CX ($\psi_\infty= -0.8\psi_{\rm cx}$) 
to realize formation of   wetting layers around the particles. 
They found that  the  spring force  
increased  above the capillary force   
transiently in an initial stage for small separation 
distances.  This effect has not yet been understood.

\begin{figure}
\begin{center}
\includegraphics[scale=0.32, angle=0]{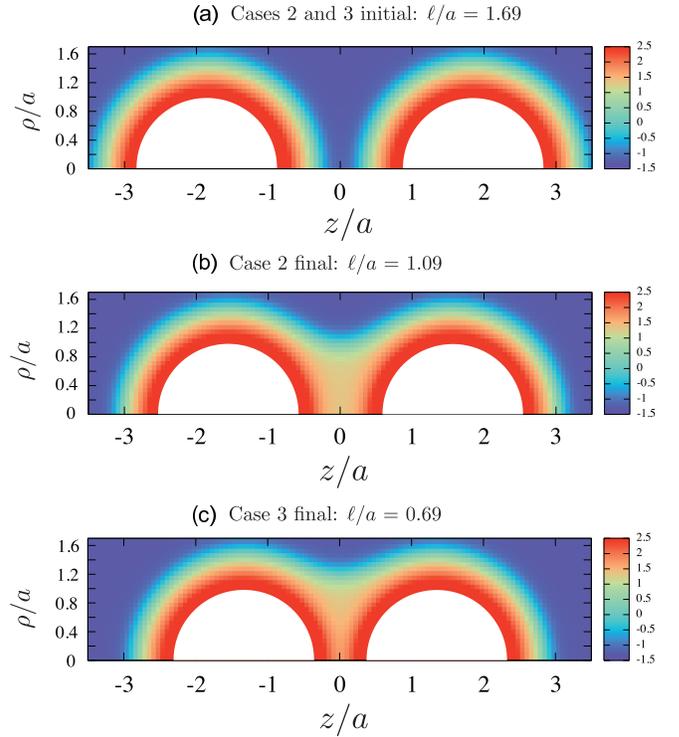}
\caption{ 
Profiles of $\psi(\rho,z)/\psi_a$ in the $\rho$-$z$ plane 
(in gradation)  with    $\tau/\tau_a=-8$  and  $\psi_\infty= 
-1.31\psi_a$ in  strong adsorption.  (a) The initial state  
for cases 2 and 3, where $\ell/a=1.69$.    (b) The final state 
    in case 2, where $\ell/a= 1.09$.   
 (c)  The final state  in case 3, 
where $\ell/a= 0.69$. 
}
\end{center}
\end{figure}

\begin{figure}
\begin{center}
\includegraphics[scale=0.2, angle=0]{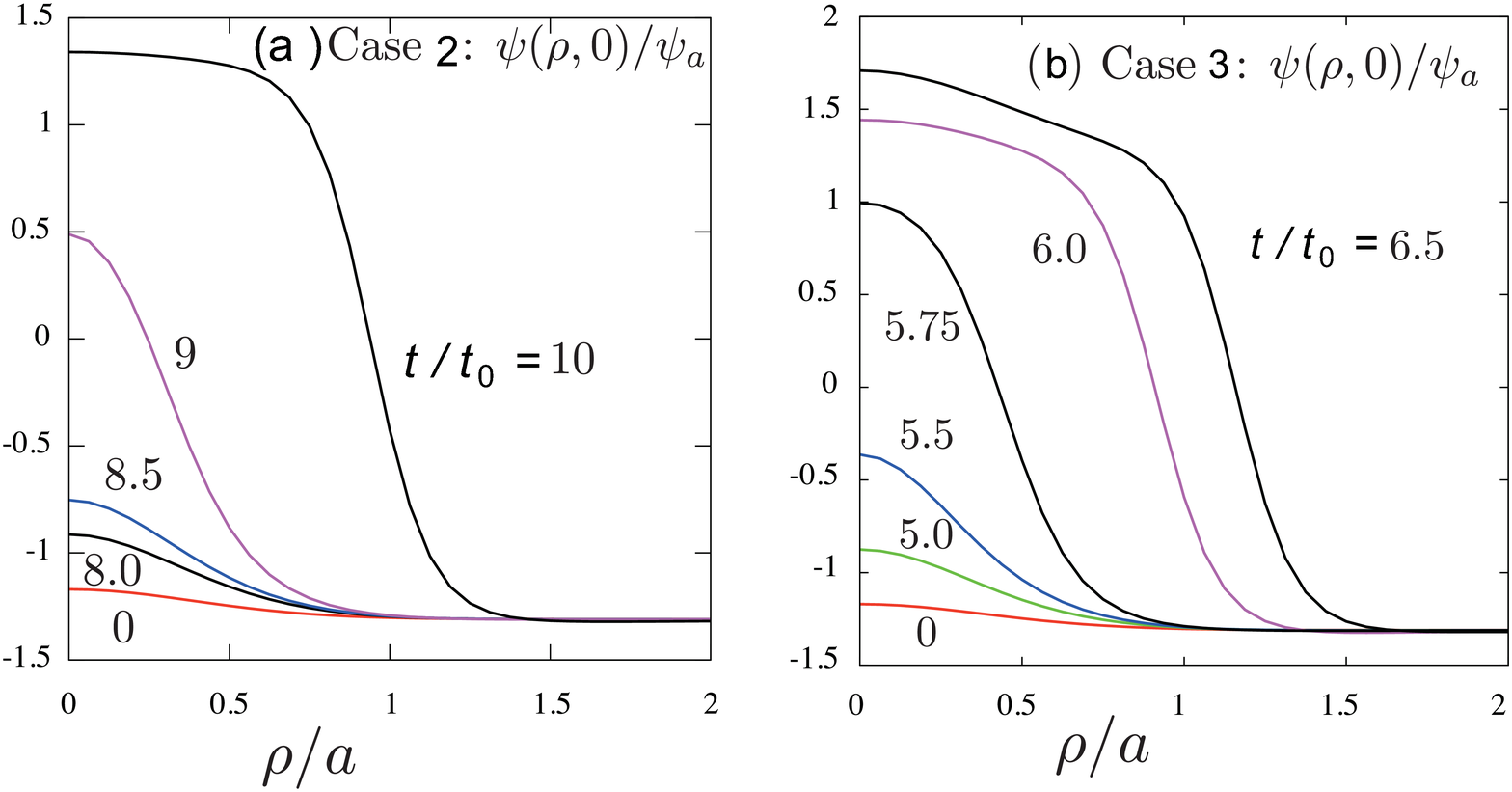}
\caption{
Time evolution of 
$\psi(\rho,0,t)/\psi_a$ on the midplane $z=0$ 
vs $\rho/a$, where 
 (a) $t/t_0=0$, 8, 8.5, 9, and  10 
 in case 2 
and  (b) $t/t_0=0$, 5, 5.5, 5.75, 6, and  6.5  
 in case 3.  
In (a), a well-defined bridging domain 
appears with a flat profile in the 
time range  $9<t/t_0<10$. 
In (b), $\psi(\rho,0,t)$ 
increases abruptly in the time range 
 $5<t/t_0<6$, where the adsorption layers of the two particles merge.  
}
\end{center}
\end{figure}

\begin{figure}
\begin{center}
\includegraphics[scale=0.2, angle=0]{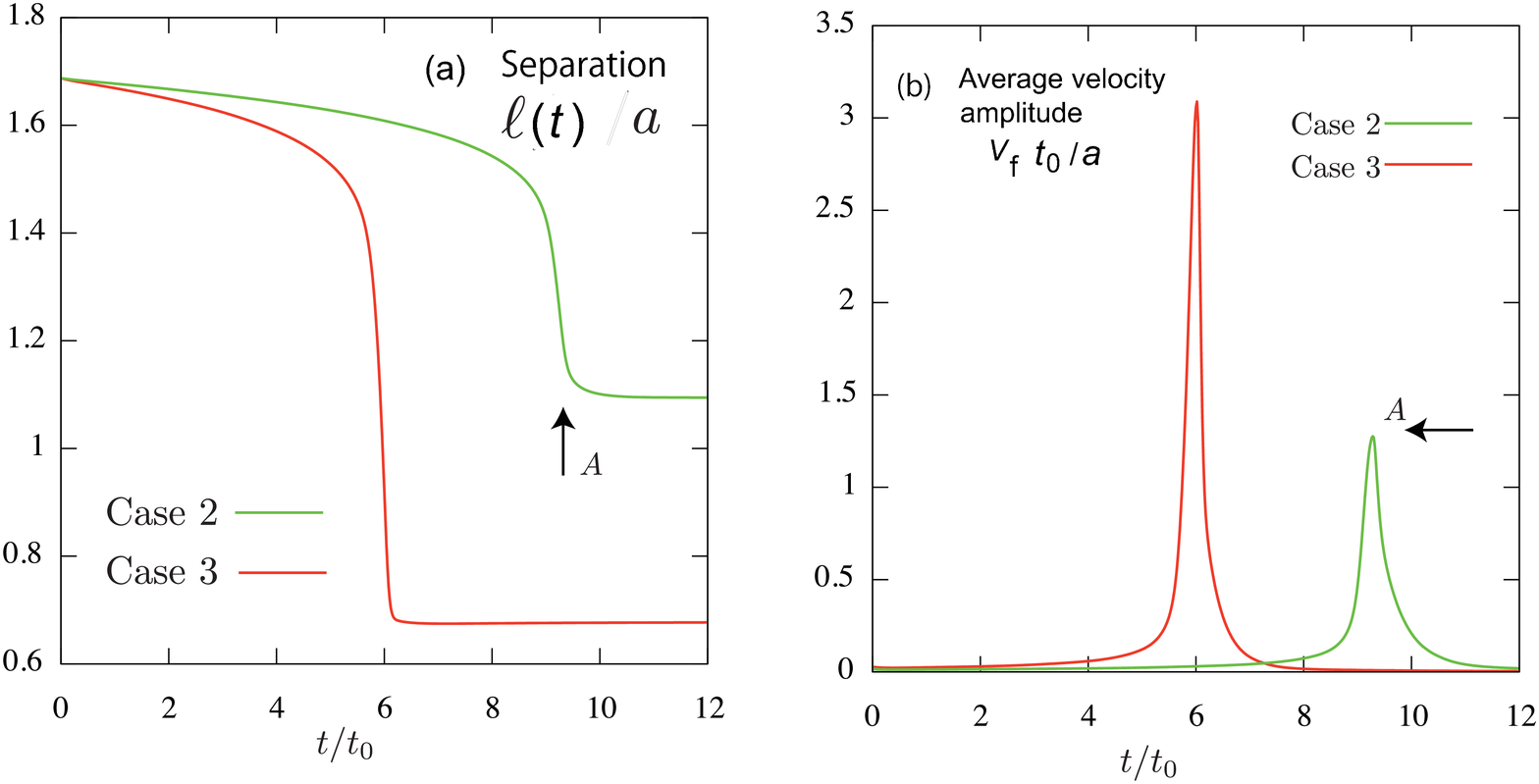}
\caption{ (a) Separation distance 
$\ell(t)$ divided by $a$ vs $t/t_0$, 
where the maximum of $|d\ell(t)/dt|$ 
is $0.457 a/t_0$ at $t=9.24t_0$ 
in case 2 and $1.27a/t_0$ at $t=6.00 t_0$  in case 3. 
The initial separation $\ell(0)$ is about  $11\xi$. 
(b) Averaged velocity amplitude $v_{\rm f}$ 
in eqn (18) divided  by  $a/t_0$  vs $t/t_0$ in cases 2 and 3.  
Peaks  are  at squeezing. 
}
\end{center}
\end{figure}

\vspace{4mm}
{\subsection{Bridging between  aggregating two particles}}

In cases 2 and 3,  we let the two particles approach 
from the initial separation  $\ell(0)= 1.69a$ 
for $t>0$, where we set  $\tau/\tau_a=-8$ and 
$\psi_\infty= -1.31\psi_a$.  Then, we have $\ell(0)/\xi\sim 11$.  
For $t<0$,   the particle positions were 
 fixed and the fluid was  in 
equilibrium, which was attained as a stationary 
solution of eqn (14).  For $t>0$, the motions are  caused by  the  
adsorption-mediated  attractive interaction.  
Notice that  the initial state is located slightly 
 below the b$\to$s instability line in Fig.1. 
This is because  the initial particle 
motions become extremely  slow above the b$\to$s line 
in the present case.  
 
The potential $U$ in eqn (4) is repulsive,  
preventing  the particles from touching each other.  
We assume that it depends   on  $\ell= 
Z_1-Z_2-2a$ exponentially as  
\be
U= k_BT_c B e^{-\kappa \ell} ,
\en
where  $B= 4\times 10^5$ in case 2 
and  $B=8\times 10^3$ in case 3 with  $\kappa =10/a$. 
As a result, we have the final separation 
distance is $\ell(\infty)= 1.09 a$ in case 2 and 
$0.69a$ in case 3.  
The above potential can arise for 
 charged colloidal particles  when  the van der Waals 
interaction is negligible\cite{Bonn}. 
For weakly charged colloid particles, 
 the screened Coulomb interaction in the 
Derjaguin, Landau, Verway,
and Overbeek (DLVO) theory \cite{Russel} is 
of the form,
\be 
U_{\rm DLVO} =
{\bar Q}^2 e^{-\kappa\ell}/[\varepsilon (1 + \kappa a)^2 
(\ell+2a)],
\en 
 where $\bar{Q}$  is the  charge on a colloid particles, 
$\epsilon$  is the 
dielectric constant, and $\kappa$ is the Debye wave number. 
If $\kappa a\gg 1$ and 
$\ell+2a\cong 2a$, $U_{\rm DLVO}$ 
assumes the form in eqn (19) 
with 
\be 
B= {\bar Q}^2 /[2k_BT_c \epsilon \kappa^2 a^3]
= 8 (\pi\rho_s/e)^2 \ell_B a /\kappa^2,
\en 
 where 
$\rho_s$ is the surface charge density 
and $\ell_B= e^2/k_BT_c\varepsilon$ is the Bjerrum length. 
For example\cite{Bonn,Die}, we may set $2\pi \rho_s/e \sim 1/$nm$^2$ 
and $\ell_B \sim 1$ nm  to obtain $B \sim 2 a^3/(a\kappa)^2$ 
with $a$ in nm, so  the adopted values of $B$ 
in this paper can be realized in experiments.

\begin{figure}
\begin{center}
\includegraphics[scale=0.3, angle=0]{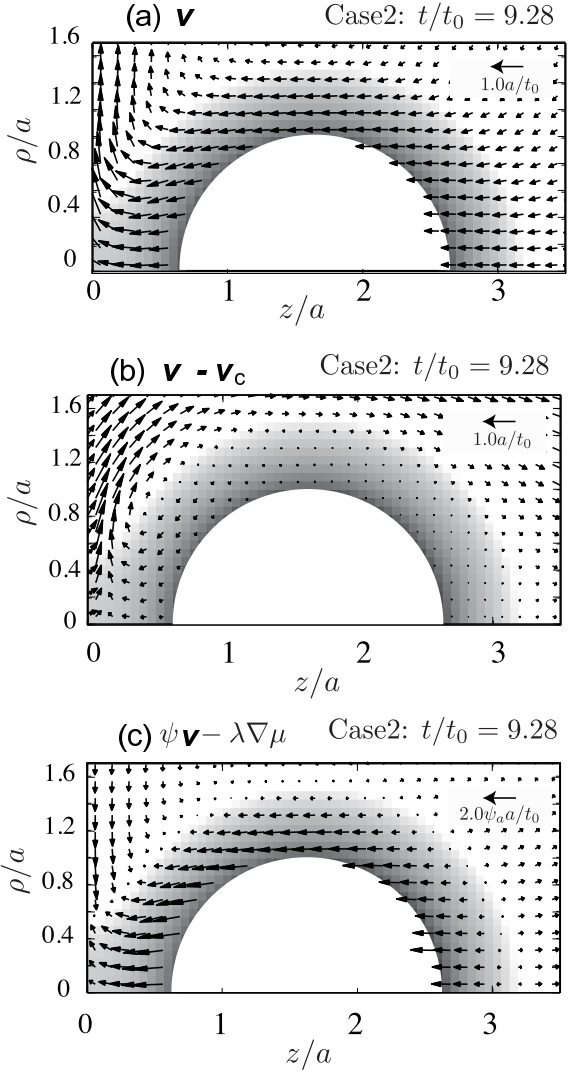}
\caption{ 
(a) Velocity field $ {\bi v}$, (b) $ {\bi v}-{\bi v}_c$,  
 and (c)  concentration  flux $\psi{\bi v}-\lambda\nabla\mu$  
at  $t/t_0=9.28$ in case 2 (point A in Fig.9), 
where ${\bi v}_c= - 0.448(a/t_0){\bi e}_z$ 
is  the particle velocity.  
Arrow lengths are written according to standard  
lengths given by    $1.0 a/t_0$ in (a) and (b) 
 and  $2.0 \psi_a a/t_0$ in (c). 
In (b), the  adsorption layer  with $\psi>0$ (in gray) moves 
with the particle far from  the midplane.
}
\end{center}
\end{figure}

\begin{figure}
\begin{center}
\includegraphics[scale=0.19, angle=0]{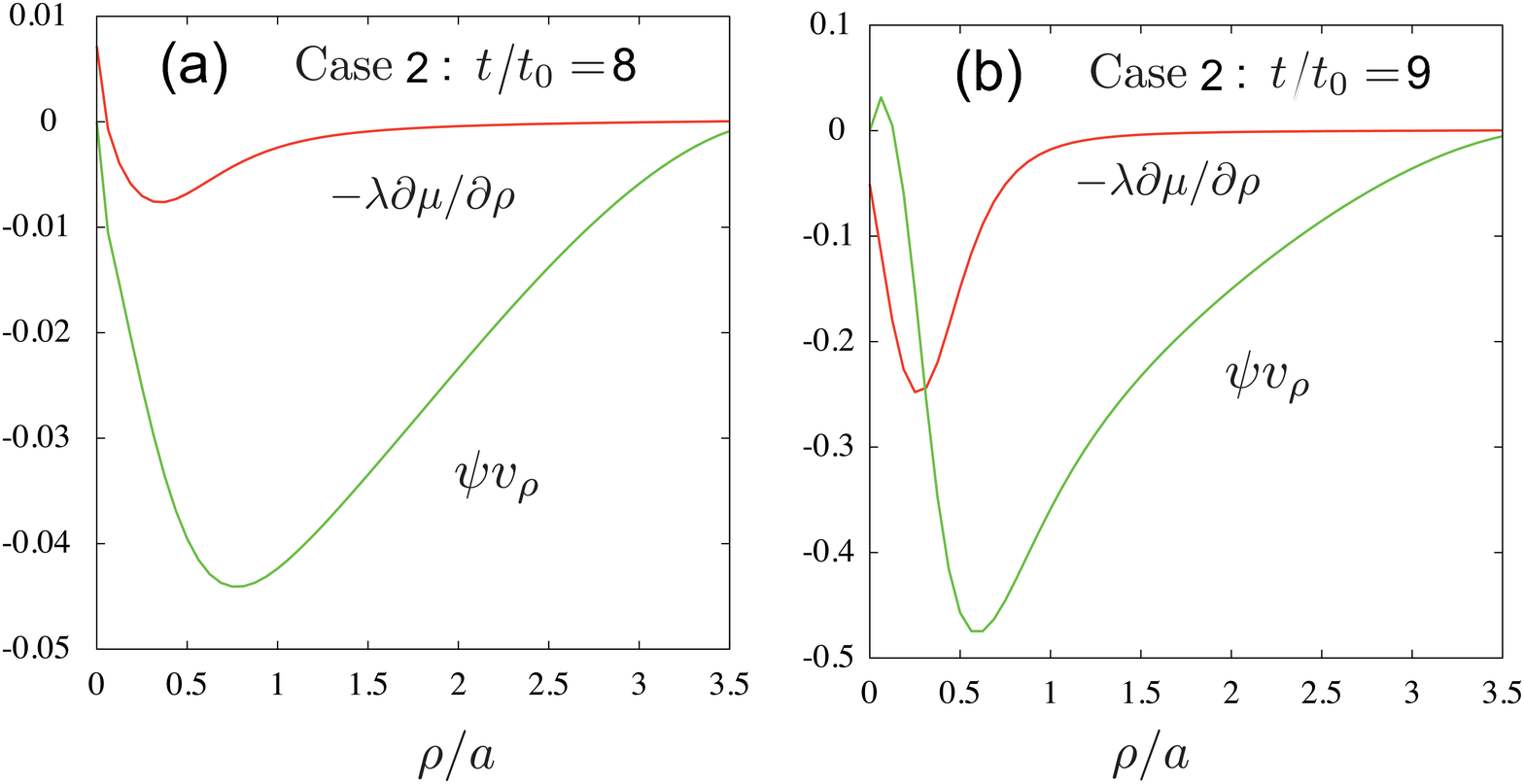}
\caption{ 
Convective flux  $\psi {v_\rho}$ 
and diffusive flux $-\lambda \p \mu/\p \rho$ vs $\rho/a$ 
in the midplane ($z=0$) 
 at (a) $t/t_0=8$  and (b) $9$  in case 2. Comparison indicates that 
the hydrodynamic convection dominates over  the  diffusive 
transport. The vertical axes are  
written in units of $\psi_a a/t_0$.  
}
\end{center}
\end{figure}

In Fig.7, we give the  profiles of 
 $\psi(r,z)/\psi_a$ in the initial and final states in cases 2 and 3. 
where  the particles are   bridged  in  the  final state. 
The adsorption layers outside the bridged region ($|z|>\ell/2$) 
 are nearly stationary as in Fig.2.  
In the final state in case 3, 
  the adsorption layers of the two particles  
are not well separated. 
In Fig.8, we show 
$\psi(\rho,0,t)/\psi_a$ on the midplane 
vs $\rho/a$  at several times. 
In case 2, $\psi(\rho,0,t)$ 
becomes fairly flat with a well-defined interface 
in the middle $\rho/a<0.7$ for $t/t_0=10$. 
In case 3,   it depends  on $\rho$ 
considerably even at the center 
for  $t/t_0=8.5$.  
The  $\psi$ between the particles 
increases abruptly in the time range $8.5<t/t_0<10$ 
in  case 2 and $5.5<t/t_0< 6.5$ in case 3. 

In Fig.9, 
we show the separation distance $\ell(t)$ 
and the average velocity amplitude $v_{\rm f}(t)$ in eqn (18). 
Here, the maximum of $|d\ell(t)/dt|$ is 
$0.457a/t_0$ at $t=9.24 t_0$ in case 2 
and is $1.27a/t_0$ at $t=6.0 t_0$ in case 3. 
The final approach in case 3 is very fast, where 
the fluid between the particles is rapidly  squeezed out.  
In Fig.10, we display (a) $ \bi {v}$, (b) $ \bi {v}-{\bi v}_c$, 
 and (c) $-\lambda \vec{\nabla} \mu+ \psi   \bi{v}$ 
at  $t/t_0=9.28$, where ${\bi v}_c= - 0.448 (a/t_0){\bi  e}_z$ 
is   the particle velocity with ${\bi e}_z$ being the 
unit vector along the $z$ axis.  The maximum velocity is 
of order $ a/t_0=  D/a$, which is  ten times larger than in Fig.5(a). 
Here, for LW, we have  
$a/t_0=1.7\times 10^{-3}$ cm$/$s for $a=200$ nm.
In (b),  the adsorption layer  moves with the particle 
and  squeezing  takes  place 
between the particles. In (c), the  
convective transport within 
the adsorption layer is crucial. 
  In Fig.11, we compare 
$\psi v_\rho$ and $-\lambda {\nabla}_\rho  \mu$ as in Fig.6, 
which demonstrates that  the 
convective transport  dominates over the diffusive transport.

It is rather surprising 
 that  the adsorption layer 
moves with the particle without large deformations  in Fig.10(b), 
though there are no 
sharp interfaces (see Fig.3(b)).  This aspect  should 
be further  examined, which  should give rise to 
a decrease  in the diffusion constant of 
suspended particles even due to critical adsorption.

\vspace{4mm}
\section{Summary and remarks}

We  have numerically examined   dynamics of bridging 
and aggregation of two colloidal particles  
in  a  near-critical  binary mixture outside CX. 
Particularly near the disfavored  branch of CX,  
the critical  adsorption is strong and extended  so that  
the adsorption-induced   interaction is 
much enhanced. 
As a result, the bridging transition readily 
takes place among the colloidal particles  when the face-to-face 
separation  $\ell$ is of the order the particle radius $ a$. 
We have assumed that $a$ is 
much longer than the correlation length $\xi$.  
We summarize our main results in the following.\\

In Sec.II, we have  explained  
the critical behavior,  the FPD method, 
  and the model H equations in the Ginzburg-Landau scheme.  
In Sec.III, we  have  explained the simulation method 
and introduced 
characteristic order parameter  $\psi_a$, reduced  
temperature $\tau_a$, and time $t_0$.  

 In Sec.IV, to  induce    a bridging transition 
in  a separated state, we have crossed the instability line 
 in the $\tau$-$\ell$ plane  by  temperature quenching (case 1)
and  by two-particle aggregation (cases 2 and 3). In  case 1,   the spring force $F_{\rm sp}(t)$ has been 
calculated as in Fig.4(a), which increases 
abruptly up to order $2\pi \sigma a$ upon 
 formation of a well-defined interface of a bridging domain. 
In cases 2 and 3, 
the separation   $\ell(t)$
has  been obtained as in Fig.9(a), where 
the two particles initially approach 
due to the adsorption-induced  attraction 
but eventually  stop due to the 
screened Coulomb interaction  in eqn (19). 
In case 2, the final distance $\ell(\infty)$ 
is  large and a well-defined  bridging domain appears. 
In case 3,  it is not large  
such  that the adsorption layers of the two particles overlap. 
In the final stage in these cases, 
the fluid between the particles is  squeezed out, 
leading to   peaks in the average velocity 
amplitude $v_{\rm f}$ in Fig.9(b).  In  all these cases, 
the convective transport of $\psi$ within the adsorption layers 
is crucial.\\

 Finally, we  further mention 
 two  experiments, which 
suggest the presence of bridging 
in  aggregation.\\
(1) 
 Broide {\it et al.}\cite{Be5} studied   aggregation kinetics 
after a temperature change at $t=0$, 
where fluidlike  aggregates 
grew  for $t>0$ with inter-particle 
separations of order   $a (\gg \xi)$. They observed that 
their  average volume $V(t)$ and   average 
length  $L(t)$   were  related by $V(t)\sim L(t)^3\propto t$  
 and  the time-dependent scattered light intensity exhibited the Porod tail\cite{Onukibook}. 
These behaviors  are very different from those 
of    fractal   aggregates induced 
 by the van der Waals interaction.\\
(2) 
 Guo {\it et al.}\cite{Guo}  observed 
a fcc crystal formed by the aggregated particles 
with $a=52.5$ nm,   where the  lattice constant was  $181$ nm 
and the colloid volume fraction was 
0.406. In their  case, the corner-to-face distance of the lattice 
was  $2.46a$ and the shortest 
face-to-face distance of the particles was $0.46a= 22$ nm. 
This separation distance 
is  much longer  than $\xi$ ($=1-10$ nm). 
We  thus propose  a bridging scenario of 
crystal formation.\\  

\noindent{\bf Acknowledgments}.
This work was supported by KAKENHI 
 (No. 25610122) and 
Grants-in-Aid for Japan Society for Promotion of 
Science (JSPS) Fellows (Grants Nos. 241799 and  263111).

%

\vspace{2mm}
\noindent{\bf Appendix: Renormalized free energy density }\\
\setcounter{equation}{0}
\renewcommand{\theequation}{A\arabic{equation}}
To  account for   the statics 
in near-critical fluids,  we use the renormalized 
local functional  theory\cite{Fisher-Yang,Up,Upton,Oka-c,Oka-b}. 
As in the linear parametric model by Schofield {\it et al.}\cite{Sch},
we introduce a distance $w$ from the critical point in the $\tau$-$\psi$ 
plane, in terms of which the overall critical behavior can be 
approximately described.  We  define  $w$   by \cite{Oka-c}
\be 
w=  \tau +C_2 w^{1-2\beta}\psi^2, 
\en  
where $C_2$ is a  constant. 
The coefficient  $C$ in eqn (4) 
is set  equal  to  $ k_BT C_1 w^{-{\hat\eta}\nu}$, 
where  $C_1$ is  a constant and $\hat\eta$ is the Fisher exponent.  

The  free energy density 
$f(\psi,\tau)$ is determined such that    the combination 
$ \xi^3[f(\psi,\tau)-f(0,\tau)]/k_BT$ is a universal function 
of $\tau/|\psi|^{1/\beta}$ in accord with the two-scale factor 
universality\cite{Onukibook,Aliu}. 
Then, we obtain the correct exponent relations from 
the free energy $F=\int d{\bi r}[f+C|\psi|^2/2]$, 
but the critical amplitude ratios \cite{Onukibook,Aliu} 
are approximate depending  on the choice of $C_2$.  
In this  paper, we use the choice  $C_2=(2\pi^2/3)C_1 \xi_0$ 
 in our previous work\cite{Oka-c}. Then, 
we obtain  $\xi_0/\xi_0'= 3.0$,  while it is 1.9 from 
  numerical analysis of the  Ising criticality \cite{Aliu}. 

In the original Schofield model\cite{Sch},
the behavior of $f$  inside CX was  not  discussed. 
We assume $f$ inside   CX (in the region 
$|\psi|<\psi_{\rm cx}$ and $\tau<0$)  in   the  form,   
\be 
{f}(\psi,\tau) ={f}_{\rm cx}+ k_B{T_c} 
(\psi^2-\psi_{\rm cx}^2)^2/
(8{\chi_{\rm cx}}{ \psi_{\rm cx}^2 })  ,
\en 
where  $f_{\rm cx}=  {f}(\psi_{\rm cx},\tau)$  and  
${\chi_{\rm cx}}=k_BT_c/(\p^2 f/\p \psi^2)$ 
on CX. Here, ${\chi_{\rm cx}}\sim 
|\tau|^{-\gamma}$  with $\gamma=(2-\hat{\eta})\nu$. 
From eqn (4) and (A2) the surface tension is calculated as  \cite{Oka-c} 
\be 
\sigma= 0.075 k_BT_c/\xi^2, 
\en 
 where $\xi=\xi_0' |\tau|^{-\nu}$ 
is the correlation length on CX. 
See also Fig.3, where 
  $\psi$ much exceeds $\psi_{\rm cx}$ 
near the colloid surfaces  and tends to a  value 
close to $-\psi_{\rm cx}$ in the bulk.   
 Thus, use of eqn (A2) is essential in our theory.

\end{document}